\documentstyle{laa}     % LaTeX A&A  Standard Fonts

\begin{document}
 
   \thesaurus{03         % A&A Section 3:Extragalactic astronomy 
	      (11.01.2;  % galaxies: active
               11.10.1;  % galaxies: jets
	       11.09.1;  % individual: 3C 279
	       13.07.3)  % gamma-rays: theory 
	       }
		
\title{On the application of the mirror model for gamma-ray flare in 3C 279}

\institute{Department of Experimental Physics, University of \L \'od\'z,
ul. Pomorska 149/153, PL 90-236 \L \'od\'z, Poland }
 
   \author{W. Bednarek }
%\offprints{Bednar@krysia.uni.lodz.pl}
 
   \date{Received  ... 1998; accepted ....  199}
 
   \maketitle
 
\begin{abstract}
Multiwavelength observations of high energy flare in 1996 from 
3C 279 seems to favour the so called mirror model between different inverse 
Compton scattering models proposed as a possible explanation of gamma-ray 
emission in blazars. We performed kinematic analysis of the relativistic blob -
mirror system and found that only part of the mirror located very close
to the jet axis (very likely inside the jet cone) can re-emit soft photons 
which serve as a target for production of $\gamma$-rays by relativistic 
electrons in the blob. Since the presence of well localized scattering mirror
inside the jet is problematic, this makes problems for the mirror model.
The time scale and the shape of the $\gamma$-ray flare should reflect, 
in terms of the mirror model, the blob dimensions and
the longitudinal distribution of relativistic electrons inside the blob. 
For the $\gamma$-ray light curve of the type observed in 1996 from 3C 279, i.e. 
the rising time of the flare during a few days with a sharp cut-off
towards the end of the flare, the density of electrons inside the blob should
increase exponentially starting from the front of the blob and reach maximum
towards the end of the blob. Such distribution of electrons is difficult to 
explain in a model of a relativistic shock moving along the jet, which would 
rather inject electrons more efficiently at the front of the blob with 
a trail of particles on its downstream side.

\keywords{galaxies: active: jets: individual: 3C 279 - gamma-rays: theory}

\end{abstract}

\section{Introduction}

About 50 blazars have been detected 
by the Compton Gamma Ray Observatory in the MeV - GeV  energy range 
(Fichtel et al. 1994, von Montigny et al. 1995, Thompson et al. 1995, 
Mukherjee et al. 1997), and 3 blazars, of the BL Lac type, are discovered 
in the TeV $\gamma$-rays by the 
Whipple Observatory (Punch et al. 1992, Quinn et al. 1996, Catanese et al. 
1997). These blazars can reach very high $\gamma$-ray luminosities which are
variable on time scales as short as a part of a day, 
in the case of optically violent variable quasars, or even several minutes, 
in the case of BL Lacs. These observations strongly suggest that 
$\gamma$-ray emission from blazars is collimated towards the observer within
a small angle as a result of relativistic motion of plasma in the jet or
directional acceleration of particles.  

High energy processes occurring in blazars are popularly explained in terms of 
the 
inverse Compton scattering (ICS) model in which $\gamma$-rays are produced in
ICS of soft photons by electrons in a blob moving 
relativistically along the jet. Different modifications of this general model
mainly concern the origin of soft photons, i.e. whether they come
internally from the blob in the jet 
(synchrotron self-Compton (SSC) model, e.g. Maraschi et al. 1992, 
Bloom \& Marscher 1993),  
directly from the disk (e.g. Dermer et al. 1992, Bednarek et al. 1996a,b),
are produced in the disk but reprocessed by the matter surrounding the disk
(external comptonization (EC) model, e.g. Sikora et al. 1994, 
Blandford \& Levinson 1995), or produced in the jet but
reprocessed by the matter surrounding the jet (the so-called mirror model, 
Ghisellini \& Madau~1996, henceforth GM). 
In this last paper it is mentioned that SSC model  and external comptonization
of photons produced by the broad line region clouds (BLR) illuminated by the 
disk (EC model) may also contribute to the $\gamma$-ray
emission producing a first $\gamma$-ray pre-flare. For the SSC model the 
amplitude of the $\gamma$-ray variation is expected to be proportional to the
square of the variation observed in IR-optical-UV energy range. For the EC
model the $\gamma$-ray emission should vary linearly with the low energy
synchrotron emission. Such behaviour is not observed in the case of the 1996
flare from 3C 279 in which the $\gamma$-ray variation is more than the square
of the synchrotron variation.
Moreover, in the $\gamma$-ray light curve of this flare (see Fig.~1 in 
Wehrle et al. 1997),
there is no clear evidence for a double peak structure which could eventually
correspond to the first $\gamma$-ray flare produced in terms of SSC or EC 
models and the
second $\gamma$-ray flare produced in terms of the mirror model.  
Therefore, although the SSC model can not be completely rule out,
Wehrle et al. (1997) concludes that the mirror model is favourite by the 
multiwavelength observations of a strong flare in February 1996 from 3C 279
since it predicts $\gamma$-ray flare with observed features.

In this paper we test the mirror model by comparing predictions of the 
kinematic analysis with the observational results. The possible
contributions from SSC and EC models to the $\gamma$-ray production during
this flare are neglected since, as we mentioned
above, there is no observational support for their importance. 
Simultaneous analysis of all these models will require an introduction of 
additional free parameters
(density of electrons in the blob, the perpendicular extend of the blob, 
definition of the disk radiation) which are not all well constrained by the 
observations.

\section{Gamma-ray production in the mirror model} 

According to Ghisellini \& Madau model, a
blob containing relativistic particles moves along the jet with the Lorentz 
factor $\gamma$ and velocity normalized to the speed of light $\beta$. 
The synchrotron radiation produced by electrons in the blob illuminates the
BLR cloud(s) (the mirror), 
located at a distance $l$ from the center of active galaxy.  
This radiation photoionizes the cloud(s) which re-emit isotropically broad
line emission. The relativistic blob approaching the mirror will 
see its radiation significantly enhanced because of decreasing blob-mirror 
distance and relativistic effects. In order to understand the features of 
this model we start this
analysis from a simple picture, i.e. a single blob with negligible longitudinal 
extend along the jet (in respect to $l$) and a one dimensional mirror located
on the jet axis. Since this picture is not successful in 
explanation of the features of $\gamma$-ray flare in 3C 279, we discuss  
more realistic case in which extended blob scatters 
radiation reflected by the two dimensional mirror.

   \begin{figure}[t]
      \vspace{8.cm}
\includegraphics{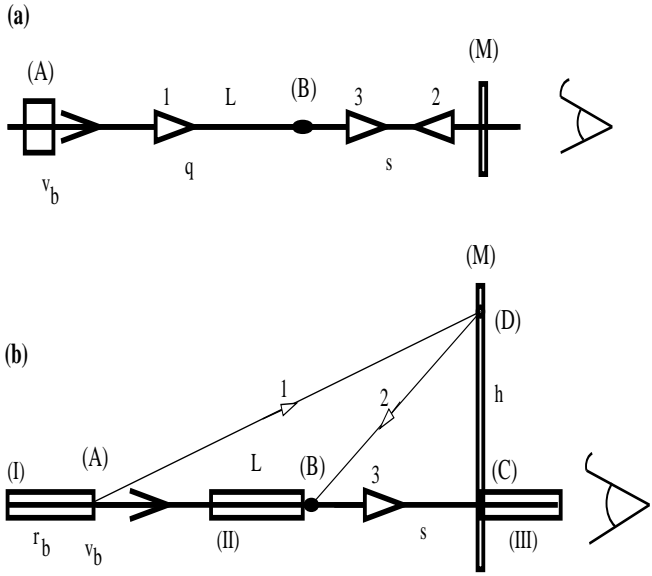}
      \caption[]{Schematic representation (not to scale) of the mirror 
model with a simple geometry, i.e. 
a single thin blob and a one dimensional mirror on the jet axis {\bf (a)}, 
and with a realistic geometry, i.e. an extended blob and a two dimensional 
plane mirror {\bf (b)}. The mirror (M) is located
at the distance $l$ from the base of the jet (A). We assume that first 
synchrotron photons (marked by 1) are produced by the blob in (A). 
First $\gamma$-rays (marked by 3) are produced by ICS of 
soft photons (marked by 2) re-emitted by the mirror, at the 
distance $s$ from the mirror (B) and the distance $q$ from the base of the jet. 
In Fig.~1b the synchrotron photons illuminate
the mirror at different locations (D) which are at a distance $h$ from the jet 
axis. Different critical locations of the blob are marked by: (I) - the blob is 
at the base of the jet; (II) - the blob starts to produce first $\gamma$-ray 
photons; (III) - the back of the blob crosses the mirror located at (C).
The blob moves with velocity $v_{\rm b}$ towards the mirror and has 
longitudinal extend $r_{\rm b}$ in case (b) and negligible
longitudinal extend in case (a) }
\label{fig1}
    \end{figure}

\subsection{A single thin blob and a one dimensional mirror}

Let us discuss the simplest possible case in which the blob, the mirror, 
and the observer are located on the jet axis (see Fig.~\ref{fig1}a).
We assume that the blob has negligible dimensions in respect to the other 
dimensions of considered system. 
First synchrotron photons are emitted by the blob at the distance marked by (A)
which has been chosen as located at the base of the jet. 
These photons (marked by 1) excite  
the mirror (marked by (M)), which is at a 
distance $l$. The photons, re-emitted by the mirror (marked by 2), 
meet the blob again at a place marked by (B) 
which is at the distance $s$ from location of the mirror. 
In (B) blob starts to produce $\gamma$-rays (marked by 3). The production of 
$\gamma$-rays stops when the blob passes through the mirror.
The path, $s$, on which $\gamma$-rays are produced by the blob, 
is given by

\begin{eqnarray}
s = \beta (1+\beta)\gamma^2c\tau_\gamma,
\label{eq1}
\end{eqnarray}  
\noindent
where $\tau_\gamma$ is the observed rise time of the $\gamma$-ray flare, and
$c$ is the velocity of light.
First $\gamma$-rays are produced at the distance $q$ from the base of the jet,
which is equal to $l - s$ and given by

\begin{eqnarray}
q = 2\beta s/(1-\beta) = 2(1+\beta)^2\beta^2 \gamma^4 c \tau_\gamma.
\label{eq2}
\end{eqnarray}

The time lag, $\tau_{\rm opt-\gamma}$, between the beginning of synchrotron 
flare, which ionizes the cloud(s), and the beginning of $\gamma$-ray flare
is 

\begin{eqnarray}
\tau_{\rm opt-\gamma} = 2(1 + \beta)\gamma^2 \tau_\gamma.
\label{eq3}
\end{eqnarray}

Eqs.~(\ref{eq2}) and ~(\ref{eq3}) show that for the case of $\gamma$-ray flares 
observed from 3C 279 (Kniffen et al. 1993, Wehrle et al. 1997) which has 
the rising time of a few days ($\tau\approx 6-8$ days in February 1996) and, 
the blob moving with typical Lorentz factor of the order of $\sim 10$ ($\gamma
> 8.5$ for 3C 279, Wehrle et al. 1997), the distance from the base of the 
jet to the place of 
$\gamma$-ray production should be of the order of a few hundred pc ($q\sim 200$
pc for 3C 279, see Eq.~(2)). The corresponding time delay between 
synchrotron and $\gamma$-ray flare should be of the order of a few years 
($\tau_{\rm opt-\gamma}\sim 2-3$ years for 3C 279).
The distance $q$ is about three orders of
magnitudes larger than the typical dimension of the BLR (see GM). 
Therefore the rise time of $\gamma$-ray flare observed in  
3C 279 can not be explained as a result of relativistic effects connected 
with the time of flight of a single thin  
blob in the radiation reflected by the BLR clouds.
However a few day time scale of the flare might be connected with
the longitudinal extension of the blob. 
Small inhomogeneities in such extended blob can be 
responsible for a short time scale variability  
of $\gamma$-ray emission (flickering)
as a result of kinematic effects discussed in this subsection.
In terms of the mirror model we can estimate the flickering time of the 
$\gamma$-ray emission during the rising time of the flare in the case of 
3C 279 by reversing Eq.~(\ref{eq2}). Assuming $q = 3\times 10^{17}$ cm and
$\gamma = 8.5$, we estimate the time scale for the shortest possible flux 
variability caused by these effects as equal to $\tau_\gamma\approx 4$ min.

\subsection{An extended blob and a two dimensional mirror}

Let us assume that the blob has longitudinal extend along the jet
$r_{\rm b}$ (see Fig.~\ref{fig1}b), and negligible perpendicular extend.  
Its perpendicular dimension do not introduce interesting effects if  
the observer is located at small angles to the jet axis.
As in the picture considered above, electrons produce
synchrotron radiation which is reflected by the mirror located at a distance
$l$ from the place of first injection of electrons 
(assumed at the base of the jet). The mirror is two-dimensional with 
negligible thickness and extends in perpendicular direction in respect 
to the jet axis. Note that Ghisellini \& Madau considered the spherical mirror.
However our assumption on the plane mirror simplifies the formulas derived 
below, because of simpler geometrical relations (rectangle triangles) and 
does not introduce any additional artifact features since only 
the part of the mirror located close to the jet axis is important. 
For simplicity we assume that the observer is located on the jet axis.
The synchrotron photons (marked by 1, in Fig.~\ref{fig1}b), 
which are produced by electrons at the place marked by (A), 
illuminate the mirror (M) at any place (D). 
The photons reprocessed by the mirror (marked by 2) meet at the first
time the blob at the distance $s$ from the mirror,
\begin{eqnarray}
s = {{1 - \beta}\over{1 + \beta}}l.
\label{eq4}
\end{eqnarray} 
\noindent
At this place first $\gamma$-rays (marked by 3) 
are produced by the blob and the $\gamma$-ray
flare begins to develop. The $\gamma$-ray emission
increases very fast up to the moment when the front of the blob meets the
mirror. This happens at the time 
\begin{eqnarray}
t_{\rm inc} = (1 - \beta)s/\beta c = l/\gamma^4 (1 + \beta)^3 \beta c, 
\label{eq5}
\end{eqnarray}
\noindent
measured from the beginning 
of the $\gamma$-ray flare. The flare finishes at the time 
\begin{eqnarray}
t_{\rm end} = t_{\rm inc} + r_{\rm b}/\beta c, 
\label{eq6}
\end{eqnarray}
when the back of the extended blob crosses the place of location of the mirror.
This equation simply relates the expected time scale of the flare to the length 
of the extended blob $r_{\rm b}$ and the distance $l$ of the mirror 
from the base of 
the jet. For relativistic blob ($\gamma \gg 1$) and the $\gamma$-ray flares 
occurring  on a time scale of days (as observed in blazars) the dependence 
of $t_{\rm end}$ on $l$ is not important (see Eqs.~(\ref{eq5}) and~(\ref{eq6})).
The full $\gamma$-ray flare is then produced on 
a distance, $r_\gamma$, measured from the mirror, which is given by
\begin{eqnarray}
r_\gamma = s + r_{\rm b}/(1 + \beta) \approx r_{\rm b}/2.
\label{eq6b}
\end{eqnarray}

Eq.~(\ref{eq6}) shows that duration of the $\gamma$-ray flare can be consistent
with the mirror model for reasonable dimensions of the blob.
However the question arises if the observed $\gamma$-ray light curves 
of flares in blazars can be
explained in such a model. Below we analyse this problem assuming different
geometries of the blob with different density distributions of relativistic
electrons.

   \begin{figure*}[t]
      \vspace{7.cm}
\includegraphics{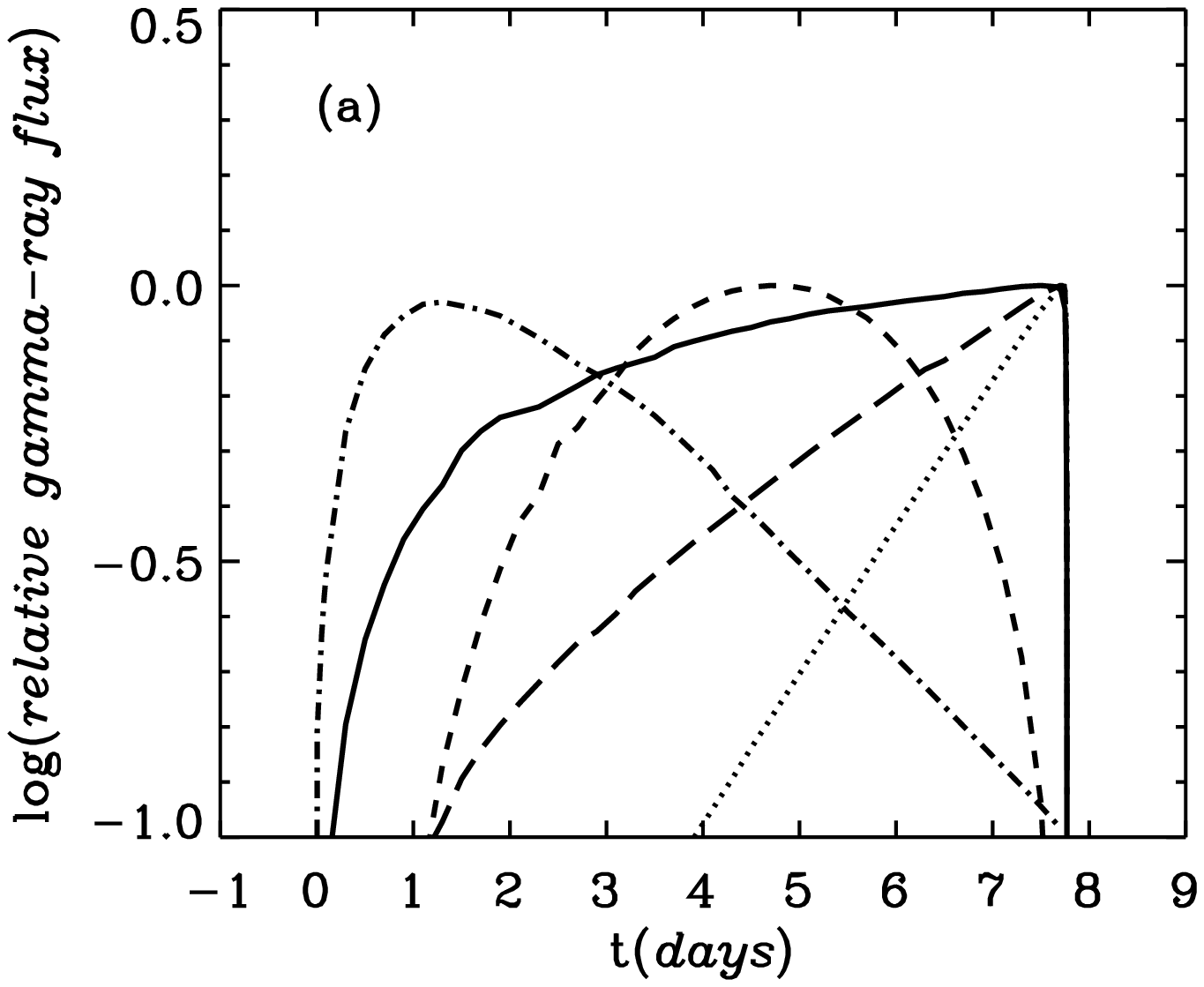}
\includegraphics{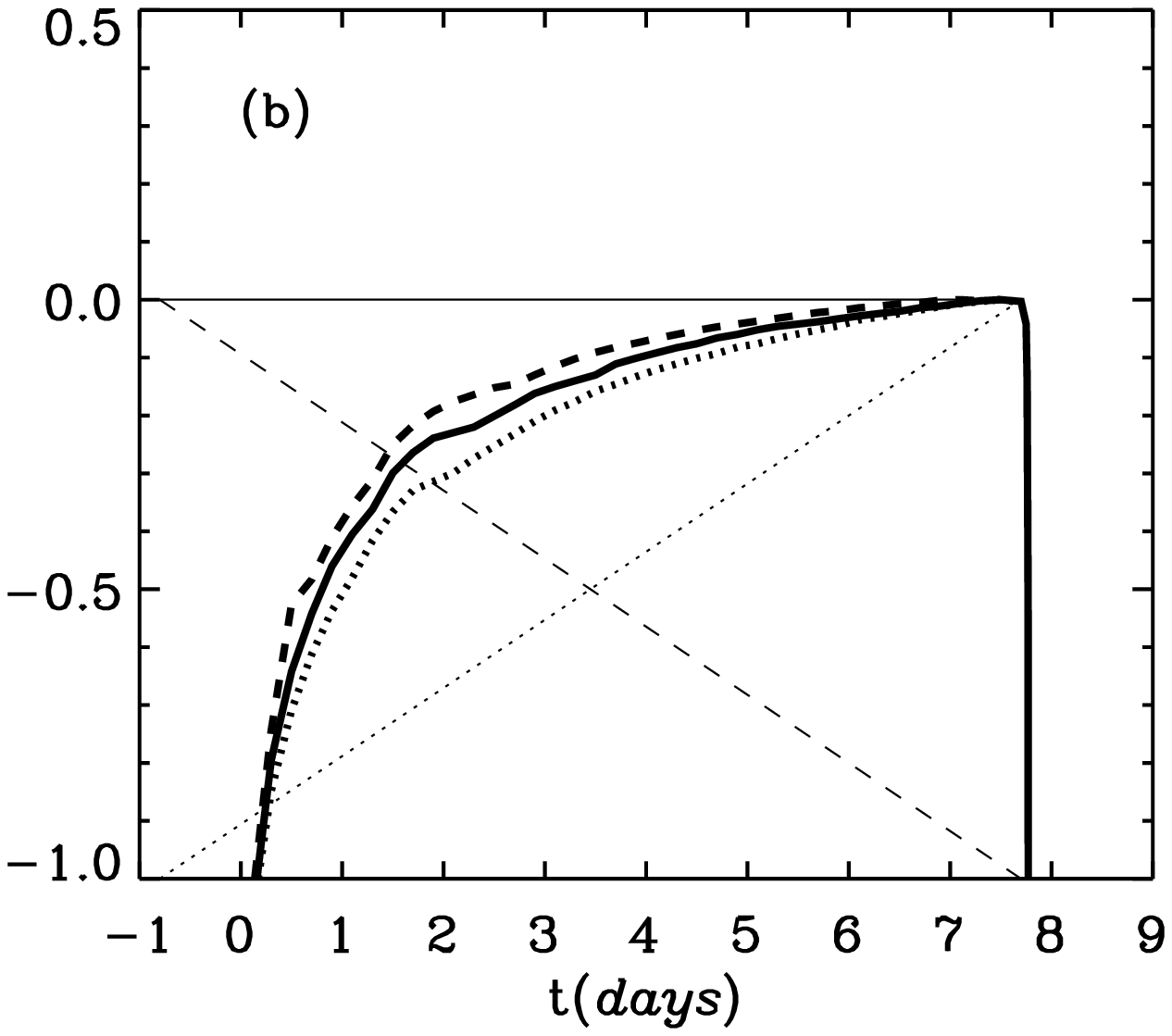}
      \caption[]{Gamma-ray light curves produced by a blob with 
different geometries and distributions of relativistic electrons. 
It is assumed that the mirror is located at the distance 
$l = 3\times 10^{17}$ cm from the place where first synchrotron photons are 
produced by the blob (close to the base of the jet). The blob has longitudinal
extend $r_{\rm b} = 2\times 10^{16}$ cm and moves along the jet with the Lorentz 
factor $\gamma = 8.5$. {\bf (a)} Different curves show the $\gamma$-ray 
light curves in the 
case of: a spherical, homogeneous blob (dashed curve), a cylindrical, 
homogeneous blob (full 
curve), and inhomogeneous blobs for the distribution of electron densities 
given by Eq.~(\ref{eq22}) (dot-dashed curve), and Eq.~(\ref{eq23}) with 
$\sigma = 10$ 
(dotted curve) and $\sigma = 2.718$ (long-dashed curve). {\bf (b)} The 
$\gamma$-ray light curves are shown by the thick curves for a cylindrical, 
homogeneous blob in which the density of 
electrons depends on the distance $x$ from the mirror according to: 
Eq.~(\ref{eq24}) (dotted curve), Eq.~(\ref{eq25}) (dashed curve) and 
$n_e = const$ (full curve). The corresponding synchrotron flares are
marked by the thin curves.}
\label{fig2}
    \end{figure*}

\subsubsection{Gamma-ray light curve produced by extended blob}

In order to determine the evolution of $\gamma$-ray power emitted 
in time $t$ by the blob, the following formula has to be integrated
\begin{eqnarray}
N_\gamma(t) = 2\pi \int_{\rm x_d}^{\rm x_u} n_\gamma n_{\rm e}(r',t') 
\int^{1}_{\mu_{m}}\int_{r_{\rm d}}^{r_{\rm u}}
I_{\rm ill}\gamma^2 (1 + \beta\mu)^2 dr d\mu dx, 
\label{eq7}
\end{eqnarray} 
\noindent
where $\mu = \cos\phi$, and $\phi$ is the angle CBD defined in Fig.~1b.
The first integral has to be performed over distances of $\gamma$-ray
emission region (part of the blob) from the mirror $x$. 
The second integral is over
different paths (defined by the cosine angle $\mu$) which has to 
be passed by synchrotron photons and
reprocessed photons in order to produce 
$\gamma$-ray photon found in time $t$ at the location of the mirror. 
This integral is equivalent to the integration
over contributions from different scattering centers of the mirror defined 
by the height 
$h$ (see Fig.~\ref{fig1}b). Therefore the limits of integration over $\mu$ can
be changed to integration over $h$ by using 
\begin{eqnarray}
d\mu = -xh(x^2 + h^2)^{-3/2}dh.
\label{eq8}
\end{eqnarray}
The third integral has to be performed over the regions in the 
blob, $r$ (measured from 
the front of the blob), emitting synchrotron photons which can produce 
reprocessed photons serving next as a target
for relativistic electrons at $x$. 

At a point defined by $h$, the mirror is illuminated by the synchrotron 
radiation from the blob with the flux (see GM)
\begin{eqnarray}
F_{\rm syn} = {{1}\over{\gamma^4(1 - \beta \cos\eta)^4}} 
{{L_{\rm syn}}\over{4\pi d^2}},
\label{eq9}
\end{eqnarray}
\noindent
where $d = (h^2 + z^2)^{1/2}$ is the distance between the regions of 
synchrotron emission and the scattering centers on the mirror, 
$z$ is the distance 
of the place of production of synchrotron photons from the mirror, and 
$\cos\eta = z/d$. The synchrotron luminosity in the blob frame can be 
expressed by 
\begin{eqnarray}
L_{\rm syn} = 4\pi r_b^2 c n_{\rm syn} n_{\rm e}(r,t''). 
\label{eq9b} 
\end{eqnarray}
\noindent
$n_{\rm syn}$ describes the synchrotron power emitted
by average relativistic electron, and $n_{\rm e}(r,t'')$ is the electron 
density in the blob as a function of $r$ at the moment $t'' = t - 
(d + \sqrt{h^2 + x^2} - x/\beta)/c$. 
For distances $d$ smaller than $r_{\rm b}$, we take in Eq.~(\ref{eq9}) 
$d = r_{\rm b}$, 
since the synchrotron luminosity can not exceed the maximum possible value 
determined by the dimension of the blob $r_{\rm b}$.

The points on the mirror at the distance, $h$, re-emits a part $a$ of 
incident flux $F_{\rm syn}$ isotropically (GM),
\begin{eqnarray}
I_{\rm ill} = {{a}\over{4\pi}}F_{\rm syn}.
\label{eq10}
\end{eqnarray} 
\noindent
The relativistic electrons with density $n_{\rm e}(r',t')$, 
responsible for production of $\gamma$-ray photons at the time $t$ 
(Eq.~\ref{eq7}), has to be counted  at the moment 
$t' = t - x/c$ and at place in the blob measured from its front 
$r' = \beta c t - (1 - \beta)(x - s)$.

The limits of integration over distances of parts of the blob from
the mirror $x$, which produce $\gamma$-rays observed at time $t$ at the 
location of the mirror, 
can be found from the analysis of propagation of the front and the back 
of the blob. The lower limit is
\begin{eqnarray}
x_{\rm d} = \cases {0, &if p $\geq$ s;\cr s-p, &if $p < s$.\cr}
\label{eq11}
\end{eqnarray}
\noindent
where $p = \beta ct/(1 - \beta)$, and the upper limit is
\begin{eqnarray}
x_{\rm u} = \cases {s + 0.5ct, &if $r_{\rm b} \geq w$;\cr s + 
(r_{\rm b} - \beta ct)/(1 - 
\beta), &if $r_{\rm b} < w$.\cr}
\label{eq12}
\end{eqnarray}
\noindent
where $w = 0.5ct(1 + \beta)$. 

Only photons reprocessed by the part of the mirror at a distance from
the jet axis smaller than $h_{\rm m}$
can contribute to the $\gamma$-ray production by the parts
of the blob located at the distance $x$ from the mirror. 
This maximum height $h_{\rm m}$ can be found by analysing
the time of flights of photons and the blob (Fig~\ref{fig1}b)
and depends on $l$ and $x$. 
It has to fulfil the following equation,
\begin{eqnarray}
l + 2s + ct = (l^2 + h_{\rm m}^2)^{1/2} + (x^2 + h_{\rm m}^2)^{1/2} +x,
\label{eq13}
\end{eqnarray}
\noindent
which has the solution 
\begin{eqnarray}
h_{\rm m} = [0.25(l + B + {{l^2 - x^2}\over{l + B}})^2 - l^2]^{1/2},
\label{eq14}
\end{eqnarray}
\noindent
where $B = 2s + ct - x$. $h_{\rm m}$ takes the maximum possible value, 
$h_{\rm u}$, if the synchrotron photons, produced on the front 
of the blob at the distance $l$ from the mirror, excite parts of the mirror 
which produce reprocessed photons 
serving next as a target for production of $\gamma$-rays by electrons 
at the moment when the back of the blob crosses the location of the mirror. 
For this constraint, following condition has to be fulfilled
\begin{eqnarray}
(l^2 + h_{\rm u}^2)^{1/2} + h_{\rm u} = (l + r_{\rm b})/\beta.
\label{eq15}
\end{eqnarray}
\noindent
The above equation has the solution
\begin{eqnarray}
h_{\rm u} = [(l + r_{\rm b})^2 -\beta^2l^2]/2\beta(l +r_{\rm b}).
\label{eq16}
\end{eqnarray}   
\noindent
For the relativistic blob ($\gamma \gg 1$) and $l/\gamma^2\ll r_{\rm b} \ll l$,
\begin{eqnarray}
h_{\rm u}\cong r_{\rm b}.
\label{eq17}
\end{eqnarray}
\noindent
Hence for the parameters of the $\gamma$-ray flares observed in 3C 279, only 
parts of the mirror close to the jet axis (laying 
mainly inside the jet) can re-emit soft photons which serve as a target
for production of $\gamma$-rays. Therefore the limits of integrations in
Eq.~(\ref{eq19}) of the paper by Ghisellini \& Madau (GM) are not correct 
because they do not take into account the dynamics of the blob.
From this reason, the energy densities of photons re-emitted by the mirror, 
but observed in the blob frame, are time independent and overestimated in 
that paper.

For given $x$ and $h$, we determine the part of the blob (its longitudinal 
extend $r$) which emits synchrotron photons. These photons initiate next 
the production of $\gamma$-ray photons observed  at the moment $t$. 
As before we analyse the time of flights of  
photons and the blob and obtain the lower limit on the longitudinal extend of the 
blob in Eq.~(\ref{eq7})
\begin{eqnarray}
r_{\rm d} = \cases {0, &if $E < m $;\cr \beta(E - h) - l, 
&if $E \geq m$.\cr}
\label{eq18}
\end{eqnarray}
\noindent
where $m = l/\beta + h$, and $E = l + B - (x^2 + h^2)^{1/2}$, and
the upper limit 
\begin{eqnarray}
r_{\rm u} = \cases {r_b, &if $r_b < r_g $;\cr r_{\rm g}, 
&if $r_{\rm b} \geq r_{\rm g}$.\cr}
\label{eq19}
\end{eqnarray}
\noindent
where $r_{\rm g} = \beta[E - (h^2 + l^2)^{1/2}]$. Finally we find the distance
$z$ of the blob from the mirror at the moment of emission of synchrotron photons
which initiate the  production of $\gamma$-ray photons at the time $t$. 
For given $x$, $h$ and $r$, we determine $z$ from the following condition
\begin{eqnarray}
F + z/\beta = (h^2 + z^2)^{1/2},
\label{eq20}
\end{eqnarray}
\noindent
with $F = E - (l + r)/\beta$. The solution of this equation is 
\begin{eqnarray}
z = (0.5\sqrt{\Delta} - \beta F)\gamma^2,
\label{eq21}
\end{eqnarray}
\noindent
where $\Delta = 4\beta^2[\beta^2F^2 + (1 - \beta^2)h^2]$.

Since we want to know the relative change of the $\gamma$-ray flux with time,
the computations of the $\gamma$-ray light curves have been 
performed assuming that the parameters describing the reflection, and
$\gamma$-ray and synchrotron efficiencies of a single
relativistic electron in the blob are $a n_\gamma n_{\rm syn} = 1$. 
In principle the values of $n_\gamma$ and $n_{\rm syn}$ may depend on the
blob propagation, e.g. if the spectrum of electrons in the blob depends on 
its propagation along the jet. We do not consider such cases in order not
to complicate the model too much. First we investigate the
dependence of the $\gamma$-ray light curve on the longitudinal distribution
of electrons in the blob, $n_{\rm e}(r)$. In general $n_e$ may depend on the
blob geometry and electron density as a function of $r$. The results of 
computations of the $\gamma$-ray light curves for a few different cases
are shown in Fig.~\ref{fig2}a. The dashed curve in this figure shows the 
$\gamma$-ray light curve in the case of cumulative distribution of electrons 
in the blob (integrated over perpendicular extend of the blob), 
$n_{\rm e}(r)\propto \sqrt{r_{\rm b}^2 - (2r - r_{\rm b})^2}$,
corresponding to the homogeneous, spherical blob with
longitudinal extend $r_b =2\times 10^{16}$ cm, moving 
with the Lorentz factor $\gamma = 8.5$. The mirror is located at the distance
$l = 3\times 10^{17}$ cm from the base of the jet. The expected light curve in 
this case is almost symmetrical with the maximum corresponding to the center
of the blob. The full curve shows the light curve for the homogeneous, 
cylindrical blob with other parameters of the model as in previous case. 
The dot-dashed curve shows the $\gamma$-ray light curve produced by the blob 
with 
cylindrical geometry but with exponential decrease of density of relativistic 
electrons. We apply the following distribution
\begin{eqnarray}
n_{\rm e}(r)\propto 10^{(r_{\rm b}-r)/r_{\rm b}}, 
\label{eq22}
\end{eqnarray}
\noindent
which might correspond to the distribution of electrons, produced by the 
relativistic plain shock, with the maximum on the front of the cylindrical 
blob and exponentially decreasing tail towards the end of the blob. 
In contrary, the density of electrons could increase exponentially with $r$, 
e.g. according to
\begin{eqnarray}
n_{\rm e}(r)\propto \sigma^{r/r_{\rm b}}. 
\label{eq23}
\end{eqnarray}
\noindent
We consider the cases with $\sigma = 10$ and $\sigma = {\rm e}\cong 2.718$, 
for which
the $\gamma$-ray light curves are shown in Fig.~\ref{fig2}a by the dotted and 
long-dashed curves, respectively. 

Fig.~\ref{fig2}b shows the $\gamma$-ray and corresponding synchrotron
light curves assuming that the density
of electrons in the blob depends on the distance $x$ from the mirror but is 
homogeneous inside the blob. These light curves are normalized to the flux at
their maximum.
The dotted curves corresponds to the continuous increase of density of 
relativistic electrons in the blob according to  
\begin{eqnarray}
n_{\rm e}(x)\propto 10^{(l-x)/l},
\label{eq24}
\end{eqnarray}
\noindent
and the dashed curves to the case when the electron density decreases according 
to
\begin{eqnarray}
n_{\rm e}(x)\propto 10^{x/l}.
\label{eq25}
\end{eqnarray}
\noindent
These $\gamma$-ray light curves are very similar to the $\gamma$-ray light 
curve (full curve) obtained in the case with constant electron density 
in the blob during its propagation in the jet.
Small differences between these $\gamma$-ray light curves are due to
the fact that the blob reaches the mirror after very short time $t_{\rm inc}$
measured from the beginning of the $\gamma$-ray flare. 
For $t > t_{\rm inc}$, the radiation field 
seen by relativistic electrons do not change significantly. Note that the
production of $\gamma$-ray photons in the blob occurs at small 
distance from the mirror (given by Eq.~(\ref{eq6b})) in comparison to the 
distance $l$ of the mirror from the base of the jet. The beginning of
the $\gamma$-ray flare is delayed in respect to the synchrotron 
flare by $t_{\rm d}= 2l/[(1+\beta)^2\gamma^2c]$. For the parameters considered in 
Fig.~2b this delay is of the order of $\sim 0.8$ day. 

In all discussed above cases the $\gamma$-ray
flux increases initially on a very short time scale (given by Eq.~(\ref{eq5})). 
For the parameters applied above this time is $t_{inc}\approx 8$ min. 
The $\gamma$-ray flare finishes at time $t_{\rm end}$ given by Eq.~(\ref{eq6}), 
which for these parameters is $\sim 7.7$ days.

\section{Confrontation with the $\gamma$-ray light curve of 1996 flare from 
3C 279}

The $\gamma$-ray blazar 3C 279 is one of the best studied up to now. 
It was the first blazar detected by the Compton GRO in June 1991, showing the 
bright flare with the $\gamma$-ray light curve which increased for a few days 
and finished on a much shorter time scale (Kniffen et al. 1993, Hartman et al. 
1996). 
Its $\gamma$-ray emission in December 1992 - January 1993 was about an order 
of magnitude lower (Maraschi et al. 1994). However on February 1996, 3C 279
again shows strong flare with the light curve very
similar to this one observed in 1991 (Wehrle et al. 1997, Collmar et al. 1997). 
Significant
variability of the $\gamma$-ray flux during this flare has been measured
on a time scale of $\sim 8$ hr. The $\gamma$-ray flux increased 
continuously for about an order
of magnitude during $6 - 8$ days and later dropped sharply during $\sim 1$ day 
(see Fig.~1 in Wehrle et al. 1997). The rapid $\gamma$-ray variability, 
simultaneous variability of X- and 10 GeV $\gamma$-rays, and the 
condition that the $\gamma$-ray emission region should be transparent, requires
for the Lorentz factor of relativistic blob $\gamma > 8.5$ (Wehrle te al. 1997).
According to Wehrle et al., the multiwavelength observations of this flare 
(the lack of evident simultaneous variations of the optical - UV flux and 
the $\gamma$-ray flux) favour 
the mirror model for the $\gamma$-ray production in this source. 

As we already mentioned in Sect.~2.1, the $\gamma$-ray flare of the type 
observed in 3C 279 can not be
produced by a single blob with negligible dimension moving with the Lorentz 
factor $\gamma = 8.5$, provided that the mirror is
located from the base of the jet at a characteristic distance of the BLR 
clouds $\sim 3\times 10^{17}$ cm. 
The rise time scale of the flare has to be connected with 
the longitudinal extend of the blob in the jet. The observed time scale
of the flare requires that the extend of the blob should be of the order of 
$r_{\rm b}\sim 2\times 10^{16}$ cm, if the mirror is at the distance 
$l = 3\times 
10^{17}$ cm and $\gamma = 8.5$. However the requirements on the time of flight
of photons and the blob show that only soft photons re-emitted by the
parts of the mirror within radius $h_{\rm u}\approx r_{\rm b}$, 
centered on the jet axis (see Eq.~(\ref{eq17})), 
can contribute to the observed $\gamma$-ray flux. Therefore the scattering
mirror has to be located within the jet cone, provided that the jet typical
opening angle is of the order of $\sim 1/\gamma$. This conclusion is 
inconsistent with the assumptions made by Ghisellini \& Madau (GM) in their
computations of the density of reprocessed photons seen by relativistic
electrons in the blob frame.

The shape of the $\gamma$-ray light curve can be explained in terms of the 
mirror model 
if the density of relativistic electrons increases exponentially
towards the end of the blob. 
Good consistency with the observed
rise time scale of the $\gamma$-ray flare in February 1996 from 3C 279 is
obtained for the density distribution of electrons in the blob of the type
$n_{\rm e}(r)\propto {\rm exp}(r/r_{\rm b})$ (see long-dashed curve in 
Fig.\ref{fig2}a) where $r$ is measured from the front of the blob.
However such distribution of electrons along the blob
is difficult to motivate in terms of the standard relativistic shock model 
moving along 
the jet. The electron distributions with the maximum on the front of the blob
and the trail streaming away from the shock on its downstream side seems to 
be more likely (Mastichiadis \& Kirk 1996, Kirk et al.~1998). 
However such electron distribution 
gives rapidly rising and exponentially decaying light curve (see 
dot-dashed curve in Fig.~\ref{fig2}a), 
which is in contrary to the observations of 3C 279. Therefore, the $\gamma$-ray
light curve of 3C 279 suggests that for the mirror model
the single large scale shock front is not likely
mechanism of injection of relativistic electrons along the blob. The sequence of
smaller scale shocks or another mechanism of acceleration of particles 
(magnetic reconnection in the jet ?) might give more appropriate explanation.

\section{Conclusion}

We discuss details of the mirror model proposed by Ghisellini \& Madau. 
This model seems to be favourite by the multiwavelength observations of the 
$\gamma$-ray flare in 1996 from 3C 279 (Wehrle et al. 1997). 
Based on the analysis of the kinematics of the emission region (a blob
moving relativistically along the jet) 
we come to the conclusion that only relatively small part of the mirror
is able to re-emit soft photons which serve as a target for production of
$\gamma$-rays. For the parameters of the $\gamma$-ray flare observed in 
1996 from 3C 279, the radius of this part of the mirror should be comparable
to the longitudinal extend of the blob. It has to be of the order of
$2 \times 10^{16}$ cm in order to be consistent with the rising time of the 
flare. This part of the mirror should lay inside the jet cone provided 
that its opening angle
is of the order of $\sim 1/\gamma$. As mentioned in Ghisellini \& Madau (GM),
the physical processes in the jet may prevent the presence of the well 
localized mirror inside the jet.

The calculations of density of photons re-emitted
by the mirror are done by Ghisellini \& 
Madau (see Fig.~2 in GM) in a time independent picture which do not 
take into account the dynamics of the blob. As a consequence they integrate
over the parts of the mirror at distances from the jet axis  which are much 
larger than the maximum distance $h_{\rm u}$ (Eq.~(\ref{eq17})), found in our 
dynamical (time dependent) analysis. The photon densities seen by the blob 
can not be directly compared with that ones obtained by us in a time dependent 
version of the mirror model. Ghisellini \& Madau results are only correct for 
the continuous (time independent)
flow of relativistic plasma along the jet axis but
overestimates the density of soft photons seen by the relativistic electrons in 
the blob with limited longitudinal extend. The relativistic blobs in blazars 
has to 
be confined to the part of the jet in order to produce the $\gamma$-ray flares
with the observed rising time scale.

We computed the $\gamma$-ray light curves expected in the dynamical version 
of the mirror model
for different distribution of relativistic electrons inside the blob
and assuming that the density of electrons in the blob changes 
during propagation along the jet. 
Slowly rising $\gamma$-ray flux with sudden cut-off towards the end of the 
flare, as observed in 3C 279,
is obtained in the case of inhomogeneous blob with electron densities 
exponentially rising towards the end of the blob. Such electron 
distribution is difficult to 
understand in the popular scenario for $\gamma$-ray production 
in which relativistic shock moves  
along the jet. It seems that such shock should rather inject 
relativistic electrons with high efficiencies
close to the front of the blob, with the trail of electrons 
on its downstream side (Kirk, Rieger \& Mastichiadis~1998). 
However the $\gamma$-ray light curve expected in this case 
is different than observed during the flares in the blazar 3C 279.

Since $\gamma$-rays are produced in a region which is close to the
mirror, therefore the shape of the $\gamma$-ray light curve is not
very sensitive on the variations of the density of 
electrons during the time of propagation of the blob between the 
base of the jet and the mirror. Of course the absolute 
$\gamma$-ray fluxes produced by the blobs with different evolutions of 
electron densities in time may differ significantly.

The $\gamma$-ray light curves presented in Figs.~\ref{fig2} show very 
sharp cut-offs towards the end of the flare due to our assumption on
the negligible thickness of the mirror. 
In fact, the observed width of the peak in the $\gamma$-ray light curve
of 3C 279, of the order of $t_m\sim 1$ day (see Fig.~1 in Wehrle et al. 1997), 
may be related to the time in which relativistic blob is moving though 
the mirror with the finite thickness. If this interpretation is correct then 
the thickness of the mirror has to be limited to 
$\rho_{\rm m}\approx c t_{\rm m} \beta (1+\beta)\gamma^2 \approx 
4\times 10^{17}$ cm
which is comparable to the distance of the mirror from the base of the jet. 

In this analysis we do not consider production of
$\gamma$-rays in terms of the SSC and EC models simultaneously 
with the mirror model since there is no
clear evidence of their importance in the $\gamma$-ray light curve
and the multiwavelength spectrum 
observed in 1996 from 3C 279 (Wehrle et al. 1997). The $\gamma$-ray light
curves reported in Figs.~2 show only relative change of the $\gamma$-ray flux 
with time. They are not straightforwardly dependent on the parameters of the 
blob (the magnetic field, electron density, blob perpendicular extend, disk 
radiation) 
which are not uniquely constrained by the observations. The SSC and EC models 
will require to fix these parameters in order to guarantee reliable 
comparisons.

\acknowledgements{This work is supported by the Polish Komitet Bada\'n 
Naukowych grant No. 2 P03D 001 14. I thank the anonymous referee for 
comments and useful suggestions.}


\begin{thebibliography}{}

\bibitem{} Bednarek, W., Kirk, J.G., Mastichiadis, A. 1996a, A\&A, 307, L17
\bibitem{} Bednarek, W., Kirk, J.G., Mastichiadis, A. 1996b, A\&AS, 120, 571
\bibitem{} Blandford, R.D., Levinson, A. 1995, ApJ, 441, 79
\bibitem{} Bloom, S.D., Marscher, A.P. 1993, in Proc. Compton Symp. eds. 
  Friedlander, M., Gehrels, N., AIP, New York, p. 578
\bibitem{} Catanese, M., Akerlof, C.W., Badran, H.M. et al. 1998, 
ApJ, in press
\bibitem{} Collmar, W., Sch\"onfelder, V., Bloemen, H. 1997, in Proc. 4th 
 Compton Symposium (Williamsburg, VA), in press (astro-ph/9711111)
\bibitem{} Dermer, C.D., Schlickeiser, R., Mastichiadis, A. 1992, A\&A, 256, 
  L27
\bibitem{} Fichtel, C.E. 1994, ApJS, 94, 551
\bibitem{} Gaidos, J.A., Akerlof, C.W., Biller, S. et al. 1996, Nat, 383, 319
\bibitem{} Ghisellini, G., Madau, P. 1996, MNRAS, 280, 67 (GM)
\bibitem{} Hartman, R.C., Webb, J.R., Marscher, A.P. et al. 1996, ApJ, 461, 698
\bibitem{} Kirk, J.G., Rieger, F.M., Mastichiadis, A. 1998, A\&A, in press 
(astr-ph/9801265)
\bibitem{} Kniffen, D.A., Bertsch, D.L., Fichtel, C.E. et al. 1993, ApJ, 411, 
   133
\bibitem{} Makherjee, R., Bertsch, D.L., Bloom, S.P. et al. 1997, ApJ, 490, 116
\bibitem{} Maraschi, L., Ghisellini, G., Celloti, A. 1992, ApJ, 397, L5
\bibitem{} Maraschi, L., Grandi, P., Urry, C.M. et al. 1994, ApJ, 435, L91
\bibitem{} Mastichiadis, A., Kirk, J.G. 1996, A\& A, 320, 19
\bibitem{} Punch, M., Akerlof, C.W., Cawley, M.F. et al. 1992, Nat, 358, 477
\bibitem{} Quinn, J., Akerlof, C.W., Biller, S. et al. 1996, ApJ, 456, L83
\bibitem{} Thompson, D.J., Bertsch, D.L., Dingus, B.L. et al. 1995, ApJS, 101, 259
\bibitem{} Sikora, M., Begelman, M.C., Rees, M.J. 1994, ApJ, 421, 153
\bibitem{} von Montigny, C. et al. 1995, ApJ, 440, 525
\bibitem{} Wehrle, A.E., Pian, E., Urry, C.M. et al. 1997, ApJ, submitted 
  (astro-ph/9711243)

\end{thebibliography}
\end{document}